\documentclass[letterpaper, 10 pt, conference]{ieeeconf}  

\IEEEoverridecommandlockouts                              
\usepackage[utf8]{inputenc}
\usepackage[T1]{fontenc}

\usepackage{booktabs}
\usepackage{bm}
\usepackage{nicematrix}
\usepackage{nicefrac}
\usepackage{makecell}
\usepackage{tikz}
\usepackage{pgfplots}
\pgfplotsset{compat=1.18}
\usetikzlibrary{arrows.meta, positioning, calc}
\usetikzlibrary{backgrounds}
\usepgfplotslibrary{patchplots}
\usepgfplotslibrary{fillbetween}
\pgfplotsset{%
    layers/standard/.define layer set={%
        background,axis background,axis grid,axis ticks,axis lines,axis tick labels,pre main,main,axis descriptions,axis foreground%
    }{
        grid style={/pgfplots/on layer=axis grid},%
        tick style={/pgfplots/on layer=axis ticks},%
        axis line style={/pgfplots/on layer=axis lines},%
        label style={/pgfplots/on layer=axis descriptions},%
        legend style={/pgfplots/on layer=axis descriptions},%
        title style={/pgfplots/on layer=axis descriptions},%
        colorbar style={/pgfplots/on layer=axis descriptions},%
        ticklabel style={/pgfplots/on layer=axis tick labels},%
        axis background@ style={/pgfplots/on layer=axis background},%
        3d box foreground style={/pgfplots/on layer=axis foreground},%
    },
}

\usepackage{siunitx}

\usepackage{amsmath}
\usepackage{amsfonts}
\usepackage{mathrsfs}
\newcommand{\End}{\hfill $\square$}

\usepackage{booktabs}
\usepackage{multirow}
\usepackage{adjustbox} 

\newtheorem{theorem}{Theorem}
\newtheorem{corollary}{Corollary}
\newtheorem{lemma}{Lemma}
\newtheorem{proposition}{Proposition}
\newtheorem{definition}{Definition}
\newtheorem{remark}{Remark}

\newtheorem{assumption}{Assumption}

\DeclareMathOperator*{\rank}{rank}

\DeclareFontFamily{U}{mathx}{\hyphenchar\font45}
\DeclareFontShape{U}{mathx}{m}{n}{
	<5> <6> <7> <8> <9> <10>
	<10.95> <12> <14.4> <17.28> <20.74> <24.88>
	mathx10
}{}
\DeclareSymbolFont{mathx}{U}{mathx}{m}{n}
\DeclareMathSymbol{\bigtimes}{1}{mathx}{"91}

\NewDocumentCommand{\tra}{om}{%
	\IfNoValueTF{#1}
	{#2}
	{#2_{[#1]}}%
}

\usepackage{tikz}
\usetikzlibrary{shadows,arrows.meta,positioning,backgrounds,fit,chains,scopes,calc,shapes}
\usetikzlibrary{backgrounds}
\definecolor{gray1}{rgb}{0.9,	0.91,	0.93}
\definecolor{lightblue}{rgb}{0.6, 0.76, 0.92}
\definecolor{gray2}{rgb}{0.95, 0.95,0.95	}
\definecolor{gray3}{rgb}{0.85, 0.85, 0.85	}
\definecolor{lightblue1}{rgb}{0.87, 0.92, 0.97	}
\definecolor{lightblue2}{rgb}{0.62, 0.76, 0.90	}
\definecolor{lightblue3}{rgb}{0.36	0.61	0.84}

\newcommand{\Tini}{{T_{\text{ini}}}}

\newcommand{\dimy}{{n_y}}

\newcommand{\dimu}{{n_u}}

\newcommand{\R}{\mathbb{R}}

\DeclareMathOperator{\symm}{symm}

\title{
Towards Passivity-informed Data-driven Predictive Control
}

\author{Guanru Pan, Oleksii Molodchyk, Mohammad Itani, Timm Faulwasser 
\thanks{The authors 
acknowledge funding by the Deutsche Forschungsgemeinschaft (DFG, German Research Foundation) -- project 519323897 and via project 520388526 (TRR 391 Spatio-temporal Statistics for the Transition of Energy and Transport).
} 
\thanks{The authors are with the Institute of Control Systems,
       Hamburg University of Technology, Hamburg, Germany
        {\tt\small $\{$guanru.pan, oleksii.molodchyk, mohammad.itani$\}$@tuhh.de, timm.faulwasser@ieee.org}}
}

\begin{document}

\maketitle
\thispagestyle{empty}
\pagestyle{empty}

\begin{abstract}
We propose a passivity-informed data-driven predictive control framework based on passivity-preserving sampling of unknown continuous-time linear systems. We incorporate the passive supply rate into the objective functional alongside an output-regulation cost, thereby explicitly accounting for energy extraction. Despite being practically motivated, the resulting objective is not always convex; its convexity depends on the collected input-output training data. We exploit passivity-induced positive semidefiniteness to derive an equivalent reformulation that remains a convex quadratic program independently of the data. Furthermore, we establish recursive feasibility and asymptotic output convergence using passivity and a terminal equality constraint. Numerical simulations illustrate the closed-loop performance and computational benefits of the proposed approach.
\end{abstract}


\section{Introduction}

Data-driven control has received increasing attention as an alternative to model-based controller design, particularly for systems for which obtaining an accurate parametric model is difficult or costly \cite{Willems2005,Coulson2019,Markovsky2021}. By directly exploiting measured input--output trajectories, data-driven methods enable prediction and control without an explicit system identification step. Nevertheless, purely data-driven approaches may neglect physical system properties. Incorporating such  information into data-driven control is therefore of particular interest, as it can improve both the interpretability and the theoretical properties of the resulting controller.

A fundamental physical property shared by many dynamical systems is dissipativity, provided that the input and output variables are chosen appropriately  \cite{Willems1972,Schaft2017}. Dissipativity characterizes the energy exchange between a system and its environment through a storage function and a supply rate. Passivity is its special case wherein the supply rate is given as $u^\top y$. Several works investigate how dissipativity and passivity properties can be characterized directly from measured data. A data-driven formulation of dissipativity based on quadratic difference forms is developed by~\cite{Maupong2017} and subsequently exploited for data-driven passivity-based control by~\cite{LorancaCoutino2022}. Data-driven verification of dissipativity properties from input--output data is considered by~\cite{Romer2019}. 

Passivity has also been exploited in model-based predictive and optimal control. In nonlinear Model Predictive Control (MPC), \cite{Raff2007,Falugi_2014,Garc_s_Ruiz_2024} use passivity to impose stability constraints rather than incorporating the physical supply rate into the objective. In contrast, \cite{Schaller2021} study energy-optimal control of port-Hamiltonian systems by minimizing the supplied power $u^\top y$. The benefits of supply-rate objectives in terms of energy efficiency and robustness are further investigated by \cite{Sen2023}. More recently, \cite{Sarkar2025} establish strict dissipativity and turnpike properties for energy-optimal control of discrete-time port-Hamiltonian systems. Extending this work to MPC, \cite{Singh2026} establish recursive feasibility and practical closed-loop stability for nonlinear discrete-time port-Hamiltonian systems. These approaches, however, rely on explicit nonlinear system models and do not address the convex formulation of data-driven predictive control with supply-rate objectives.

We propose a passivity-informed data-driven predictive control framework for unknown linear time-invariant (LTI) systems.
The main contributions are threefold:
i) We combine passivity-preserving sampling \cite{Stramigioli2005} with Willems' fundamental lemma \cite{Willems2005} to obtain an exact, passivity-preserving data-driven representation.
ii) We formulate a predictive controller with a supply-rate objective that accounts for energy extraction and establish recursive feasibility and asymptotic output convergence using passivity and a terminal equality constraint.
iii) Even in the LTI setting, stage costs constructed  from passive supply rates lead to indefinite, thus nonconvex, quadratic objective functionals. Exploiting the passivity-induced positive semidefiniteness property of \cite{Romer2019}, we derive an equivalent convex quadratic program (QP), independently of the collected input-output data.
Numerical simulations demonstrate the closed-loop performance and computational benefits compared with the original nonconvex formulation and conventional quadratic-cost data-driven predictive control.

The remainder of this paper is organized as follows. Section~\ref{sec:pre} introduces the passivity-preserving sampling scheme and the data-driven system representation. Section~\ref{sec:predictive_control} presents the passivity-informed predictive control framework, its convex reformulation, and closed-loop guarantees, along with a practical extension to noisy data. Section~\ref{sec:simulation} illustrates the proposed approach through numerical simulations, and Section~\ref{sec:con} concludes the paper.

\section{Preliminaries}~\label{sec:pre}


Consider an \emph{unknown} passive continuous-time  LTI system
\begin{subequations}\label{eq:PH}
\vspace{-5mm}
\begin{align}
    \dot{x} &= Ax+Bu,\quad x(0) = x_\text{ini}\\
    y &= Cx + Du,
\end{align}
\end{subequations}
with the dimension of the input variable, denoted by $\dimu$ being equal to that of the output one, denoted by $\dimy$, and the dimension of the state variable being denoted by $n_x$. 
Passivity of \eqref{eq:PH} implies that there exists a quadratic
storage function
\begin{equation}
    S(x)=\frac{1}{2}x^\top Px,\qquad P\succeq0,
\end{equation}
such that for any $t_0\leq t_1$,
\begin{equation}\label{eq:passivity}
    S(x(t))\bigg\rvert_{t_0}^{t_1}
    \leq \int_{t_0}^{t_1}u(t)^\top y(t)\,\mathrm{d}t.
\end{equation}

Since the system matrices $A$, $B$, $C$, and $D$ are unknown, neither the state $x$ nor the storage function $S$ (nor the dissipated power) is assumed to be available for controller design. Instead, we rely on measured input-output data and the prior knowledge that system~\eqref{eq:PH} is passive.

To construct a data-driven representation of~\eqref{eq:PH}, the continuous-time input and output have to be sampled. However, standard sampling schemes do not, in general, preserve passivity; see~\cite{Jiang1993}. We thus employ the passivity-preserving  scheme by~\cite{Stramigioli2005}.

We consider piecewise-constant inputs with sampling period $T_s \in \mathbb R_{>0}$, i.e., the input is held constant on every sampling interval
\begin{subequations}\label{eq:sampling}
\begin{equation}\label{eq:piece_wise_u}
    u(t)=u_k,\qquad
    t\in[kT_s,(k+1)T_s), \,\, k\in \mathbb Z_{\geq0}.
\end{equation}
 We define the sampled output as the average
\begin{equation}\label{eq:sampled_y}
    y_k
    \doteq
    \frac{1}{T_s}
    \int_{kT_s}^{(k+1)T_s}y(t)\,\mathrm{d}t.
\end{equation}
The discrete-time state is
\begin{equation}
x_k \doteq x(kT_s) .
\end{equation}
\end{subequations}
Then, system~\eqref{eq:PH} is discretized as
\begin{subequations}\label{eq:discrete_LTI}
\begin{align}
x_{k+1}&=A_\mathrm{d}x_k+B_\mathrm{d}u_k,\quad x_0 = x_\text{ini} \label{eq:discrete_state}\\
y_k&=C_\mathrm{d}x_k+D_\mathrm{d}u_k, \label{eq:discrete_output}
\end{align}
\end{subequations}
where
\begin{align*}
A_\mathrm{d}&\doteq e^{AT_s},\,\,
B_\mathrm{d}\doteq\int_0^{T_s}e^{A\tau}B\,\mathrm{d}\tau,\,\,
C_\mathrm{d}
\doteq
\frac{1}{T_s}
\int_0^{T_s}
Ce^{A\tau}\,\mathrm{d}\tau,\\
D_\mathrm{d}
&\doteq D+
\frac{1}{T_s}
\int_0^{T_s}
\int_0^\tau
Ce^{As}B\,\mathrm{d}s\,\mathrm{d}\tau.
\end{align*}
Since the continuous-time system matrices $A$, $B$, and $C$ are unknown, their discrete-time counterparts $A_\mathrm{d}$, $B_\mathrm{d}$, $C_\mathrm{d}$, and $D_\mathrm{d}$ are also unknown.
Hence, the representation~\eqref{eq:discrete_LTI} is used solely for analysis and not for control design.

Under the piecewise-constant input~\eqref{eq:piece_wise_u} and averaged output~\eqref{eq:sampled_y}, the supplied energy over each sampling interval satisfies
\begin{equation*}
\begin{aligned}
\int_{kT_s}^{(k+1)T_s}u^\top(t)y(t)\,\mathrm{d}t
&=u_k^\top\int_{kT_s}^{(k+1)T_s}y(t)\,\mathrm{d}t=T_su_k^\top y_k.
\end{aligned}
\end{equation*}
In contrast, pointwise output sampling, e.g.,
$y_k=y(kT_s)$, does not generally preserve the
supplied energy.

The following lemma shows that the passivity of the continuous-time system~\eqref{eq:PH} is preserved in the sampled system~\eqref{eq:discrete_LTI} in the discrete-time sense~\cite[Section IV.A]{Stramigioli2005}.
\begin{lemma}
\label{lem:sampled_passivity}
Consider the continuous-time system~\eqref{eq:PH} under the sampling strategy~\eqref{eq:sampling}. The corresponding discrete-time system~\eqref{eq:discrete_LTI} is passive, i.e.,
\begin{equation}\label{eq:discrete_passivity}
    S_\mathrm{d}(x_{k+1})-S_\mathrm{d}(x_k)
    \leq u_k^\top y_k, \,S_\mathrm{d}(x)\doteq S(x)/T_s.
\end{equation}
\end{lemma}
\begin{remark}
If the averaged output~\eqref{eq:sampled_y} is not directly measurable, it can be approximated using measurements at a finer sampling period $T_s'=T_s/N_s$, with $N_s\in\mathbb Z_{>1}$:
\begin{equation}\label{eq:sampled_y_approx}
    y_k
    \approx
    \frac{1}{N_s+1}
    \sum_{\ell=0}^{N_s}
    y(kT_s+\ell T_s').
\end{equation}
In the following, we assume that the exact averaged output~\eqref{eq:sampled_y} is available.
\End
\end{remark}

We next establish a data-driven representation of the sampled system~\eqref{eq:discrete_LTI} using Willems' fundamental lemma. To this end, we impose the following  assumption.

\begin{assumption}\label{ass:cont_det}
The pair $(A_\mathrm{d},B_\mathrm{d})$ is controllable.
\End
\end{assumption}

Assumption~\ref{ass:cont_det} is mild if the continuous-time pair $(A,B)$ is controllable. In such case, the sampled pair $(A_\mathrm{d},B_\mathrm{d})$ remains controllable except for pathological sampling periods; see Chapter~6.7 of~\cite{Chen1999}.

Under Assumption~\ref{ass:cont_det}, Willems' fundamental lemma provides a data-driven representation of the sampled system, provided that the input data are persistently exciting~\cite{Willems2005}. We first recall the definition of persistent excitation; we denote by $v_{[0,T-1]}$ the restriction of a signal $v$ to the interval $\mathbb{Z}_{[0,T-1]}\doteq\mathbb{Z}\cap [0,T-1]$.

\begin{definition}[Persistency of excitation]
A signal $v_{[0,T-1]}\in\mathbb R^{n_v T}$ is said to be \emph{persistently exciting of order $L\in\mathbb Z_{>0}$} if $T\in \mathbb Z_{\geq(n_v+1)L-1}$ and the Hankel matrix
\[
\mathcal H_L(v_{[0,T-1]})
\doteq
\begin{bmatrix}
v_0 & \cdots & v_{T-L}\\
\vdots & \ddots & \vdots\\
v_{L-1} & \cdots & v_{T-1}
\end{bmatrix}
\]
has full row rank.
\End
\end{definition}

\begin{lemma}[Willems' fundamental lemma]\label{lem:fundamental}
Consider system~\eqref{eq:discrete_LTI} and Assumption~\ref{ass:cont_det}, and let $(u^\mathrm{d},y^\mathrm{d})$ be a trajectory of length $T\in\mathbb Z_{>0}$. Suppose that $u^\mathrm{d}$ is persistently exciting of order $L+n_x$. Then, $(u_{[0,L-1]},y_{[0,L-1]})$ is a length-$L$ trajectory of~\eqref{eq:discrete_LTI} if and only if there exists $g\in\mathbb R^{T-L+1}$ such that
\begin{equation}
\begin{bmatrix}
\mathcal H_L(u^\mathrm{d})\\
\mathcal H_L(y^\mathrm{d})
\end{bmatrix}
g
=
\begin{bmatrix}
u_{[0,L-1]}\\
y_{[0,L-1]}
\end{bmatrix}. \tag*{\End}
\end{equation}
\end{lemma}
\section{Passivity-Informed Data-Driven Predictive Control}~\label{sec:predictive_control}

To formulate a data-driven predictor for the optimal control problem,  we partition the Hankel matrices into past and future components as
\begin{equation}\label{eq:Hankel_partitioning}
\mathcal H_{\Tini+N}(u^\text d)
=
\begin{bmatrix}
U_\text p\\
U_\text f
\end{bmatrix},
\qquad
\mathcal H_{\Tini+N}(y^\text d)
=
\begin{bmatrix}
Y_\text p\\
Y_\text f
\end{bmatrix},
\end{equation}
where, with $n_g \doteq T-\Tini-N+1$, $U_\text p \in \mathbb \R^{\Tini n_u \times n_g}$ and $Y_\text p \in \mathbb \R^{\Tini n_y \times  n_g}$ contain the first $\Tini\in\mathbb Z_{>0}$ block rows, while $U_\text f \in \mathbb \R^{N n_u \times  n_g}$ and $Y_\text f \in \mathbb \R^{N n_y \times  n_g}$ contain the remaining $N \in \mathbb Z_{\geq \Tini}$ block rows. 
At time instant $k\geq\Tini$, the measured past input-output trajectory is
\[
u_{\text p,k}\doteq u_{[k-\Tini,k-1]}\in\mathbb R^{\dimu\Tini},
\,
y_{\text p,k}\doteq y_{[k-\Tini,k-1]}\in\mathbb R^{\dimy\Tini}.
\]
The predicted future trajectories are denoted by
\[
u_{\text f,k}\doteq u_{[0,N-1]|k},
\qquad
y_{\text f,k}\doteq y_{[0,N-1]|k},
\]
where $u_{i|k}$ and $y_{i|k}$ denote the predicted input and output at time $k+i$, respectively.
Let $Q\in \R^{n_y\times n_y}$, with $Q=Q^\top$ and $Q\succeq 0$. Using Lemma~\ref{lem:fundamental}, we formulate the following data-driven optimal control problem. 
\begin{subequations}\label{eq:OCP}
\begin{align}
J_k^\star\doteq
\min_{g}
&\sum_{i=0}^{N-1}
u_{i|k}^\top y_{i|k}
+
\|y_{i|k}\|_Q^2
\label{eq:objective}\\
\text{s.t.}\quad \begin{bmatrix}
U_\text p\\
Y_\text p
\end{bmatrix}g&
=
\begin{bmatrix}
u_{\text p,k}\\
y_{\text p,k}
\end{bmatrix},
\label{eq:deepc_consistency}\\
\begin{bmatrix}
U_\text f\\
Y_\text f
\end{bmatrix}g&
=
\begin{bmatrix}
u_{\text f,k}\\
y_{\text f,k}
\end{bmatrix},
\label{eq:behavioral}\\
u_{i|k}\in\mathbb U,&\qquad
y_{i|k}\in\mathbb Y,
\quad
\forall i\in\mathbb Z_{[0,N-\Tini-1]},
\label{eq:input_output}\\
u_{i|k}=0,&\qquad
y_{i|k}=0,\quad \forall i \in \mathbb Z_{[N-\Tini,N-1]}
\label{eq:terminal}
\end{align}
\end{subequations}

Here, the objective~\eqref{eq:objective} combines the passive supply rate $u^\top y$, which represents power exchange with the environment, with the quadratic output penalty $\|y\|_Q^2$ for output regulation. Constraint \eqref{eq:deepc_consistency} ensures that the input--output trajectory predicted in \eqref{eq:behavioral} is consistent with the past measurements.
To establish closed-loop guarantees, we impose the terminal equality constraint~\eqref{eq:terminal} following~\cite{Berberich2020}, which enables a shifted optimal trajectory to serve as a feasible candidate at the next sampling instant. Additionally, \eqref{eq:input_output} constrains the remaining predicted input-output trajectory to the sets $\mathbb U$ and $\mathbb Y$, both containing the origin in their interiors. In the following we denote by $\mathbb{U}^N$ the set consisting of all sequences $u_{[i,i+N-1]}$ such that $u_k\in\mathbb{U}$, $k\in \mathbb{Z}_{[i,i+N-1]}$; $\mathbb{Y}^N$ is defined analogously. 

The supply-rate term $u^\top y$, together with the terminal equality constraint \eqref{eq:terminal}, implicitly accounts for energy dissipation without requiring state measurements. Indeed, the energy balance of the passive system yields
\[
E_{\mathrm{diss}}
= T_s\sum_{i=0}^{N-1}u_{i|k}^\top y_{i|k}
+S(x_k),
\]
where  the predicted terminal state is zero. Since $x_k$ is fixed by the measured past, minimizing the accumulated supply rate is equivalent to minimizing the dissipated energy.

Unlike conventional data-drviven predictive control with convex quadratic tracking costs, the supply-rate objective in \eqref{eq:OCP} generally leads to a nonconvex optimization problem, complicating its numerical solution. We next show how passivity can be exploited to recover convexity.

\subsection{Convex reformulation of the OCP}

Substituting $u_{\mathrm f,k}=U_{\mathrm f}g$ and $y_{\mathrm f,k}=Y_{\mathrm f}g$ into \eqref{eq:objective} yields the quadratic objective
\[
g^\top\bigl(
\operatorname{symm}(U_{\mathrm f}^\top Y_{\mathrm f})
+Y_{\mathrm f}^\top\bar QY_{\mathrm f}
\bigr)g,
\]
where $\bar Q=I_N\otimes Q$ and
$\operatorname{symm}(A)=(A+A^\top)/2$.

The Hessian of this quadratic objective is not necessarily positive semidefinite, confirming that the OCP \eqref{eq:OCP} is generally nonconvex. Nevertheless, we show that passivity guarantees convexity when the problem is expressed in suitable reduced coordinates.
To this end, we impose the standard assumption on the initialization horizon.
\begin{assumption}[Initialization horizon]\label{ass:Tini}
There exists $\bm\ell \in \mathbb Z_{\geq0}$, $\bm\ell\leq\Tini$ such that
$
\operatorname{rank}\mathcal O_{\bm\ell}(A_\mathrm d,C_\mathrm d)=n_x,
$
with
$
\mathcal O_{\bm\ell}(A_\mathrm d,C_\mathrm d)
\doteq
\begin{bmatrix}
C_\mathrm d^\top &
A_\mathrm d^\top C_\mathrm d^\top &
\cdots &
(A_\mathrm d^{\bm\ell-1})^\top C_\mathrm d^\top
\end{bmatrix}^\top. $ \End
\end{assumption}

Under Assumption~\ref{ass:Tini}, the past input-output trajectory $(u_{\text p,k},y_{\text p,k})$ uniquely determines the initial state $x_k$ of the prediction horizon~\cite{Markovsky2008}.

We introduce a reduced-coordinate parameterization of  $g$ that satisfies the past consistency condition~\eqref{eq:deepc_consistency}. Define
\[
g_0\doteq
\begin{bmatrix}U_\text p\\Y_\text p\end{bmatrix}^{\dagger}
\begin{bmatrix}u_{\text p,k}\\y_{\text p,k}\end{bmatrix},
\quad
r_\text p\doteq
\rank\begin{bmatrix}U_\text p\\Y_\text p\end{bmatrix}.
\]
Let $N_\text p\in\mathbb R^{n_g\times n_\eta}$ be an orthonormal basis of $\ker([U_\text p^\top\;Y_\text p^\top]^\top)$, where $n_\eta\doteq n_g-r_\text p$.
Then, every $g$ satisfying~\eqref{eq:deepc_consistency} admits the representation
\begin{equation}\label{eq:column_selector}
g=g_0+N_\text p\eta,
\qquad \eta\in\mathbb R^{n_\eta}.
\end{equation}
Consequently, the predicted future trajectories in~\eqref{eq:behavioral} are parameterized as 
\begin{equation}\label{eq:reduced_prediction}
u_{\text f,k}=U_\text f(g_0+N_\text p\eta),\quad
y_{\text f,k}=Y_\text f(g_0+N_\text p\eta).
\end{equation}

Since $[U_\text p^\top\;Y_\text p^\top]^\top N_\text p=0$, the free component $N_\text p\eta$ generates a trajectory with zero past input-output data. By Assumption~\ref{ass:Tini}, the corresponding future trajectory
\[
u_\eta\doteq U_\text fN_\text p\eta,
\qquad
y_\eta\doteq Y_\text fN_\text p\eta
\]
starts from the zero state. Passivity therefore implies
\begin{equation}\label{eq:nonnegative_supply}
\begin{aligned}
\eta^\top N_\text p^\top U_\text f^\top Y_\text fN_\text p\eta
&=\sum_{i=0}^{N-1}u_{\eta,i}^\top y_{\eta,i} \geq0,\quad \forall\eta\in\mathbb R^{n_\eta}.
\end{aligned}
\end{equation}
This yields the following positive-semidefiniteness property, which is a special case of the data-driven dissipativity characterization in~\cite[Theorem~2]{Romer2019}. 

\begin{corollary}[Passivity-induced positive semidefiniteness]
\label{cor:passivity_psd}
Suppose Assumptions~\ref{ass:cont_det}--\ref{ass:Tini} hold and let $N_\text p$  be an
orthonormal basis of
$\ker([U_\text p^\top\;Y_\text p^\top]^\top)$. Then,
\begin{equation*}
M
\doteq
N_\text p^\top
\symm\left(
U_\text f^\top Y_\text f
\right)
N_\text p = M^\top
\succeq0. \tag*{$\hfill \square$}
\end{equation*}
\end{corollary}

We  now exploit this property to establish an equivalent convex reformulation of~\eqref{eq:OCP}.

Using the reduced-coordinate parameterization~\eqref{eq:column_selector}, and re-using $\overline Q\doteq I_N\otimes Q$, we get
\begin{align}
H
&\doteq
M
+
N_\text p^\top
Y_\text f^\top\overline QY_\text f
N_\text p,
\label{eq:H}\\
h_k
&\doteq
N_\text p^\top
\left(
\symm(U_\text f^\top Y_\text f)
+
Y_\text f^\top\overline QY_\text f
\right)g_0,
\label{eq:h}\\
c_k
&\doteq
g_0^\top
\left(
\symm(U_\text f^\top Y_\text f)
+
Y_\text f^\top\overline QY_\text f
\right)g_0.
\label{eq:c}
\end{align}
To express the terminal equality constraint~\eqref{eq:terminal} in reduced coordinates, we further partition
\[
U_\text f=
\begin{bmatrix}
U_{\text f,1}\\
U_{\text f,2}
\end{bmatrix},
\qquad
Y_\text f=
\begin{bmatrix}
Y_{\text f,1}\\
Y_{\text f,2}
\end{bmatrix},
\]
where $U_{\text f,2}$ and $Y_{\text f,2}$ contain the last $\Tini$ block rows of $U_\text f$ and $Y_\text f$, respectively.
Now  the  reduced-coordinate optimization problem reads
\begin{subequations}\label{eq:convex_OCP}
\begin{align}
\widehat J_k^\star
=
\min_{\eta}\quad&
\eta^\top H\eta+2h_k^\top\eta+c_k
\label{eq:convex_objective}\\
\text{s.t.}\quad&
U_{\text f,1}(g_0+N_\text p\eta)\in\mathbb U^{N-\Tini},
\label{eq:convex_input}\\
&
Y_{\text f,1}(g_0+N_\text p\eta)\in\mathbb Y^{N-\Tini},
\label{eq:convex_output}\\
&
U_{\text f,2}(g_0+N_\text p\eta)=0,
\\
&Y_{\text f,2}(g_0+N_\text p\eta)=0.
\label{eq:convex_terminal}
\end{align}
\end{subequations}

The following result establishes that~\eqref{eq:convex_OCP} is a convex quadratic program and is equivalent to the original data-driven optimal control problem~\eqref{eq:OCP}.

\begin{proposition}[Equivalent convex reformulation]
\label{pro:convex_reformulation}
Let $Q\succeq0$ and suppose that the conditions of
Corollary~\ref{cor:passivity_psd} hold. Then 
\eqref{eq:convex_OCP} is a convex quadratic program equivalent
to~\eqref{eq:OCP}. In particular, the optimal values coincide, i.e.
$
\widehat J_k^\star=J_k^\star,
$
and the optimal predicted input-output trajectories obtained from the
two formulations coincide. \End
\end{proposition}

\begin{proof}
Substituting \eqref{eq:reduced_prediction} into the
objective of \eqref{eq:OCP} yields
\[
J_k(\eta)=\eta^\top H\eta+2h_k^\top\eta+c_k.
\]
By Corollary~1 and $Q\succeq0$, we have $H\succeq0$ due to \eqref{eq:H}.
The parameterization preserves the feasible predicted
input-output trajectories and their objective values.
Hence, \eqref{eq:OCP} and its reduced formulation
are equivalent, and the latter is a convex quadratic
program.
\end{proof}
Related model-based convex reformulations of indefinite quadratic MPC are developed in \cite{Zanon2014}. In contrast, our approach exploits passivity directly in behavioral coordinates without requiring a system model or storage function.

\subsection{Closed-loop guarantees}

Following the dissipativity-based stability analysis of economic MPC \cite{Angeli2012,Faulwasser2018}, we exploit passivity and the terminal equality constraint to establish recursive feasibility and asymptotic convergence of the closed-loop output.

\begin{theorem}[Recursive feasibility and output convergence]
\label{thm:closed_loop}
Suppose Assumptions~\ref{ass:cont_det}--\ref{ass:Tini} hold, that $Q\succ0$, and that the OCP~\eqref{eq:OCP} is feasible at some time instant $k\geq\Tini$. Then, the OCP remains feasible for all $j\geq k$. Moreover, the resulting closed-loop output satisfies
$
\lim_{j\to\infty}y_j=0. $ \End
\end{theorem}

\begin{proof}
\textit{Recursive feasibility.}
Let $\{u_{i|k}^\star,y_{i|k}^\star\}_{i=0}^{N-1}$ denote an optimal input-output trajectory at time instant $k$. Consider the shifted candidate trajectory at time instant $k+1$ 
\begin{subequations}\label{eq:shifted_trajectory}
\begin{align}
&\tilde u_{i|k+1}=u_{i+1|k}^\star,\quad
\tilde y_{i|k+1}=y_{i+1|k}^\star,
&&i\in\mathbb Z_{[0,N-2]},\\
&\tilde u_{N-1|k+1}=0,\quad
\tilde y_{N-1|k+1}=0. \label{eq:appended_terminal}
\end{align}
\end{subequations}
Since the last $\Tini$ input-output samples of the optimal trajectory at time $k$ are zero by~\eqref{eq:terminal}, the shifted trajectory again contains $\Tini$ terminal zero input-output samples. Moreover, the first $\Tini$ samples of the shifted trajectory are consistent with the measured past at time $k+1$. Hence, by the fundamental lemma, the shifted trajectory satisfies the behavioral constraint~\eqref{eq:behavioral}, the input-output constraints~\eqref{eq:input_output}, as well as the terminal equality constraint~\eqref{eq:terminal}. Thus, feasibility at time $k$ implies feasibility at time $k+1$, and recursive feasibility follows by induction.

\textit{Output convergence.}
Define the rotated stage cost
\begin{equation*}
\ell(u_{i|k},y_{i|k})
\doteq
u_{i|k}^\top y_{i|k}
+\|y_{i|k}\|_Q^2
+S_\text d(x_{i|k})-S_\text d(x_{i+1|k}).
\end{equation*}
By the discrete-time passivity inequality~\eqref{eq:discrete_passivity},
\begin{equation}\label{eq:rotated_stage_bound}
\ell(u_{i|k},y_{i|k})
\geq
\|y_{i|k}\|_Q^2
\geq0.
\end{equation}
Consider the rotated optimal value
\begin{equation*}
V_k^\star
\doteq
\sum_{i=0}^{N-1}
\ell(u_{i|k}^\star,y_{i|k}^\star) =
J_k^\star
+S_\text d(x_{0|k}^\star)-S_\text d(x_{N|k}^\star).
\end{equation*}

By the terminal equality constraint and Assumption~\ref{ass:Tini}, the terminal zero input-output trajectory corresponds to the zero final state, i.e.  $x_{N|k}^\star=0$. Since $x_{0|k}^\star=x_k$ and $S_\text d(0)=0$, we obtain
\begin{equation}\label{eq:rotated_value_J}
V_k^\star
=
J_k^\star+S_\text d(x_k).
\end{equation}
Furthermore, \eqref{eq:rotated_stage_bound} implies $V_k^\star\geq0$.

Evaluating the rotated value along the shifted feasible trajectory~\eqref{eq:shifted_trajectory} gives
\begin{align*}
\tilde V_{k+1}
&\doteq
\sum_{i=0}^{N-1}
\ell(\tilde u_{i|k+1},\tilde y_{i|k+1})=
\sum_{i=1}^{N-1}
\ell(u_{i|k}^\star,y_{i|k}^\star).
\end{align*}
where the appended input and output pair~\eqref{eq:appended_terminal} contributes zero. Consequently,
\begin{equation*}
\tilde V_{k+1}-V_k^\star
=
-\ell(u_{0|k}^\star,y_{0|k}^\star)
\leq
-\|y_{0|k}^\star\|_Q^2.
\end{equation*}
Since the shifted trajectory is feasible at time $k+1$, optimality gives $J_{k+1}^\star\leq\tilde J_{k+1}$. Moreover, the initial state at time $k+1$ coincides with the state reached after applying the first optimal input at time $k$, i.e., $x_{0|k+1}^\star=x_{1|k}^\star=x_{k+1}$. Using~\eqref{eq:rotated_value_J} for the optimal and shifted trajectories therefore yields $V_{k+1}^\star\leq\tilde V_{k+1}$. Hence,
\begin{equation}\label{eq:rotated_value_decrease}
V_{k+1}^\star-V_k^\star
\leq
-\|y_k\|_Q^2,
\end{equation}
where $y_k=y_{0|k}^\star$ under the receding-horizon implementation.
Summing~\eqref{eq:rotated_value_decrease} for all $j\geq k$ and using $V_j^\star\geq0$ yields
\[
\lim_{K\to \infty}\sum_{j=k}^{K}\|y_j\|_Q^2\leq V_k^\star<\infty.
\]
Since $Q\succ0$, it follows that $\displaystyle \lim_{j\to\infty}y_j=0$, which proves the claimed output convergence. 
\end{proof}

\subsection{Practical formulation for noisy data}
In the presence of measurement noise, the passivity-induced positive semidefiniteness in Corollary~\ref{cor:passivity_psd} may be violated:  the matrix $\widehat M$ constructed from noisy data may have negative eigenvalues. To retain convexity while accounting for these  passivity violations, define
\begin{equation*}
\widehat M
\doteq
N_\text p^\top
\symm\left(
U_\text f^\top Y_\text f
\right)N_\text p
=
V\Lambda V^\top,
\end{equation*}
where $V$ is orthogonal and $\Lambda=\operatorname{diag}(\sigma_1,\ldots,\sigma_{n_\eta})$ contains the eigenvalues of $\widehat M$. We decompose $\widehat M=\widehat M_+-\widehat M_-$, where
\begin{align*}
\widehat M_+&
\doteq V\operatorname{diag}(\max\{\sigma_i,0\})V^\top \succeq 0,
\\
\widehat M_-&
\doteq V\operatorname{diag}(\max\{-\sigma_i,0\})V^\top \succeq 0.
\end{align*}
 We replace the nominal quadratic cost matrix by
\begin{equation}\label{eq:H_practical}
H_\text{prac}
\doteq
\widehat M_+
+
N_\text p^\top Y_\text f^\top\overline QY_\text fN_\text p
+
\lambda\widehat M_-,
\end{equation}
where $\lambda \in \mathbb R_{>0}$ is a user-defined regularization parameter. The affine term $h_k$ in~\eqref{eq:h} is retained. This yields the practical OCP
\begin{subequations}\label{eq:practical_OCP}
\begin{align}
J_{k,\text{prac}}^\star
\doteq
\min_{\eta}\quad&
\eta^\top H_\text{prac}\eta+2h_k^\top\eta+c_k
\label{eq:practical_objective}\\
\text{s.t.}\quad&
\eqref{eq:convex_input}-\eqref{eq:convex_terminal}.
\end{align}
\end{subequations}
Since $H_\text{prac}\succeq 0$, the formulation~\eqref{eq:practical_OCP} thus provides a convex implementation for noisy data by retaining the passivity-consistent positive-semidefinite component while regularizing directions associated with  passivity violations. A rigorous  analysis of robustness and  closed-loop guarantees under noisy data are beyond the scope of this paper and are left for future work.

\section{Simulation}~\label{sec:simulation}
To investigate the performance of our OCP formulations, we use a benchmark model of a brushed DC motor rotating an inert load via a flexible shaft. The ground-truth parameters of this system are listed in Table~\ref{tab:parameters}; its continuous-time state-space model matrices (with $n_x=4$, $n_u = n_y = 2$) read
\begin{equation*}
\resizebox{\columnwidth}{!}{$
\begin{aligned}
    A &= \begin{bmatrix}
        -R_a/L_a & -K_e/L_a & 0 & 0 \\
        K_t/J_m & -(B_m+d)/J_m & d/J_m & -k/J_m \\
        0 & d/J_L & -(B_L+d)/J_L & k/J_L \\
        0 & 1 & -1 & 0
    \end{bmatrix}, \\
    B &= \begin{bmatrix}
        1/L_a & 0 & 0 & 0 \\
        0 & 0 & 1/J_L & 0
    \end{bmatrix}^{\top}, \,
    C = \begin{bmatrix}
        1 & 0 & 0 & 0 \\
        0 & 0 & 1 & 0
    \end{bmatrix}, \,
    D = 0_{2\times 2}.
\end{aligned}
$}
\end{equation*}
The state vector comprises $x = [i_a \; \omega_m \; \omega_L \; \phi]^\top$, where $i_a$ (in \si{\ampere}) is the motor armature current, $\phi$ (in \si{\radian}) is the shaft twist angle, whereas $\omega_m$ and $\omega_L$ (in \si{\radian/\second}) are the angular velocities of the rotor and load, respectively. The two inputs $u = [V_a \;\; \tau_L]^\top$ are the motor armature voltage $V_a$ (in \si{\volt}) and the brake torque $\tau_L$ (in \si{\newton\metre}). Notice that the smallest integer $\bm \ell$ satisfying the rank condition in Assumption~\ref{ass:Tini} is equal to $\bm \ell = 2$. Hence, to this end we set $T_\mathrm{ini} = 2$ as the length of the initialization horizon and we state all quantities in SI units.
\begin{table}
    \centering
    \caption{Ground-truth system parameters}
    \label{tab:parameters}
    \begin{tabular}{lcSl}
        \toprule
        Armature resistance    & $R_a$   & 0.06    & \si{\ohm} \\
        Armature inductance    & $L_a$   & 0.05    & \si{\milli\henry} \\
        Torque constant        & $K_t$   & 0.045   & \si{\newton\metre\per\ampere} \\
        Back-EMF constant      & $K_e$   & 0.045   & \si{\volt\second\per\radian} \\
        \midrule
        Rotor inertia          & $J_m$   & 0.003   & \si{\kilo\gram\metre\squared} \\
        Rotor viscous friction & $B_m$   & 0.006   & \si{\newton\metre\second\per\radian} \\
        \midrule
        Load inertia           & $J_L$ & 0.1    & \si{\kilo\gram\metre\squared} \\
        Load viscous friction  & $B_L$ & 0.002   & \si{\newton\metre\second\per\radian} \\
        \midrule
        Shaft stiffness    & $k$     & 0.5       & \si{\newton\metre\per\radian} \\
        Shaft damping      & $d$     & 0.006   & \si{\newton\metre\second\per\radian} \\
        \bottomrule
    \end{tabular}
\end{table}

We draw inspiration from regenerative braking. In this setting one is interested in controlling the motor until it is fully stopped. This task has to be performed by withdrawing \emph{exclusively} the electrical energy out of the system, i.e., without acting by applying braking pads and converting kinetic energy into heat. As for the electrical port, both the voltage $V_a$ and the current $i_a$ are usually restricted due to safety considerations. This motivates setting the constraint sets to $\mathbb U = [-48, 48] \times \{0\}$ and $\mathbb Y = [-100, 100] \times \mathbb R$, where we leave the angular velocity unconstrained.

Simulations are performed in Julia on a PC with a $1.3~\si{\giga\hertz}$ Intel i5-1335U CPU and $16~\si{\giga\byte}$ RAM. Convex  QP are solved using \texttt{DAQP.jl} \cite{arnstrom2022dual}, whereas for nonconvex problems (i.e., \eqref{eq:OCP} for the case when $
\symm(U_\mathrm{f}^\top Y_\mathrm{f}) + Y_\mathrm{f}^\top\overline{Q} Y_\mathrm{f} \succeq 0
$ does not hold), we use \texttt{Ipopt.jl} with \texttt{JuMP.jl} \cite{ipopt}. For all our tests, we set the sampling time to $T_s = 50~\si{\milli\second}$ and work with a prediction horizon of $N=400$. We generate the data for the Hankel matrices by setting the system state to $[0\;\; 300\;\;300\;\; 0]^\top$ and exciting the system by applying the piecewise constant input voltage $V_a = 13.5 + V_{a,\mathrm{rand}}$ to the motor, wherein $V_{a,\mathrm{rand}}$ is drawn randomly with uniform probability from the set $\{-5,-4,\ldots,4,5\}$ at every time step. The input-output trajectory is collected for $T=1000$ time steps (50 seconds).  
Since the input torque $\tau_L$ is restricted to zero, we do not apply it during training data collection. This allows us to halve the size of the input Hankel matrices. To avoid dimensionality mismatch, in the product $U_\mathrm{f}^\top Y_\mathrm{f}$ we only multiply voltages $V_a$ with currents $i_a$. Thus, we focus on the electrical energy supplied to the system over the prediction horizon. In the term $Y_\mathrm{f}^\top\overline{Q} Y_\mathrm{f}$ related to braking, we keep both outputs and we set $Q\doteq \operatorname{diag}([0.8\;\;5]) \cdot 10^{-6}$ to prioritize the decrease in $\omega_L$ over $i_a$.

\subsection{Noise-Free Case}
First, we study the closed-loop trajectories under ideal conditions, where the output average samples \eqref{eq:sampled_y} are exact and noise-free. To avoid computing averages at every time step, here we simulate the system response using the discretized model \eqref{eq:discrete_LTI}.

Plugging the data collected under these conditions into the Hankel matrices, we observe that the smallest eigenvalue of the Hessian $\operatorname{symm}(U_{\mathrm f}^\top Y_{\mathrm f})
+Y_{\mathrm f}^\top\bar QY_{\mathrm f}$ of \eqref{eq:OCP} is $\approx-17.676$.
This renders \eqref{eq:OCP} nonconvex, which makes a case for the equivalent convexified formulation in \eqref{eq:convex_OCP}.

For comparison, we produce closed-loop simulations using receding-horizon controllers based on: i) OCP \eqref{eq:OCP} with $g$ as decision variable, ii) convexified OCP \eqref{eq:convex_OCP} in the reduced coordinates $g_0 \oplus \operatorname{im}(N_\mathrm{p})$ \eqref{eq:column_selector}; and iii) \eqref{eq:convex_OCP} rewritten with $U_\mathrm{f}^\top U_\mathrm{f}$ in the matrix \eqref{eq:H} to use the term $u^\top u$ in the stage cost instead of the passivity-related supply rate $u^\top y$.

\begin{figure}[t]
    \centering
    \input{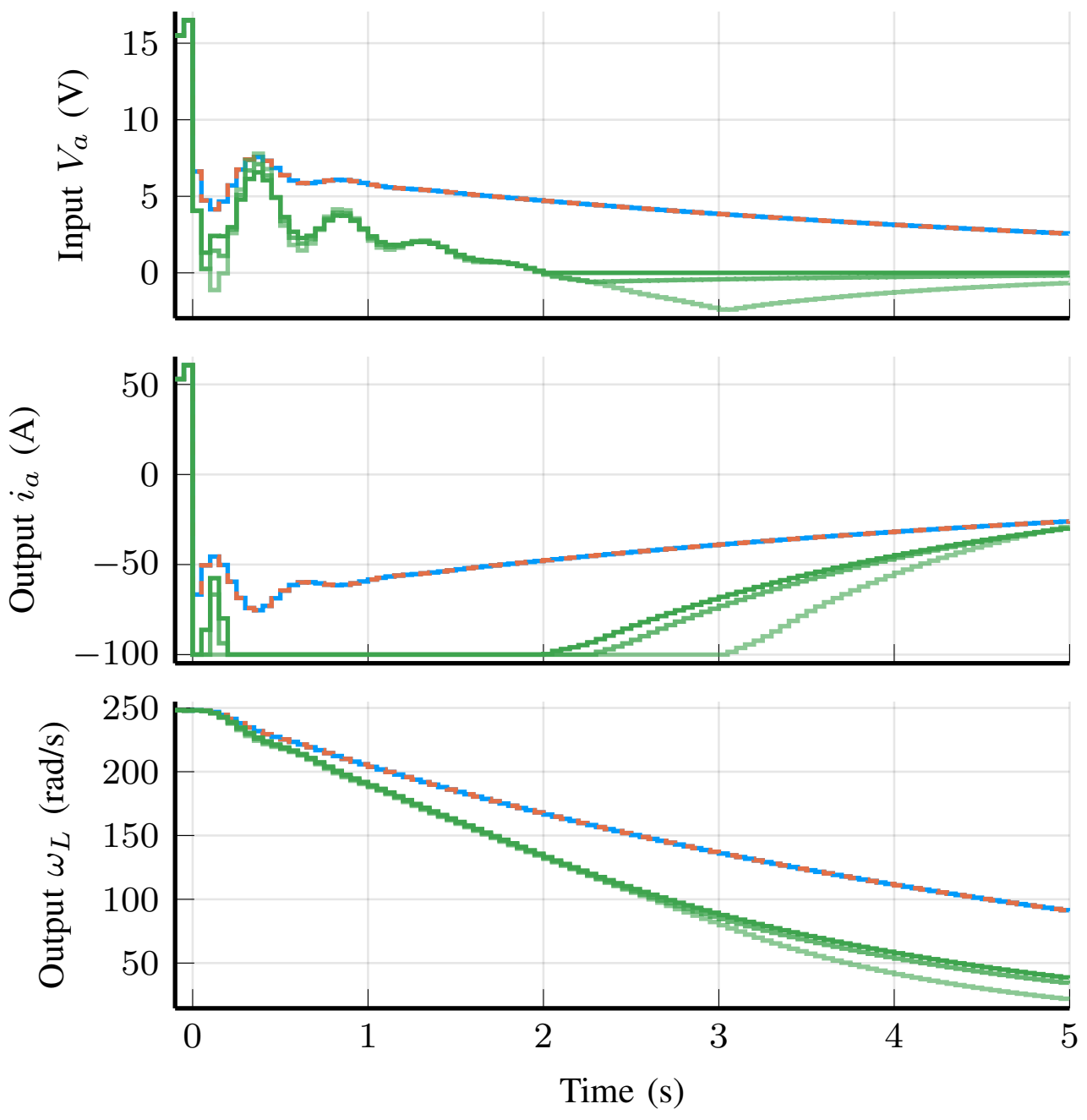}%
    \caption{Closed-loop input $V_a$ and outputs $i_a$ and $\omega_L$ for controllers with different OCP formulations.
Blue solid: stage cost $u^\top y + y^\top Q y$ (non-convex QP \eqref{eq:OCP}); orange dashed: stage cost $u^\top y + y^\top Q y$ (convex QP \eqref{eq:convex_OCP});
green: stage cost $\alpha \cdot u^\top u + y^\top Q y$ (convex QP), with shading from light ($\alpha=0.1$) to medium ($\alpha=1$) to dark ($\alpha=1$, $Q = 0$).}
    \label{fig:u_y_no_noise}
\end{figure}

Figure~\ref{fig:u_y_no_noise} shows the difference between the control decisions taken using $u^\top y$ vs. $u^\top u$ as stage cost terms. The controller guided by the minimization of the input magnitudes naturally keeps the voltage closer to zero. This, however, induces large currents in the motor's armature, immediately hitting the constraint of $-100~\si{\ampere}$ and causing the controller to produce voltage oscillations to maintain the current within the prescribed bounds. We observe this behavior across a wide range of weightings used for the term $u^\top u$. On the other hand, using the term $u^\top y$ that incentivizes the extraction of electrical energy over the prediction horizon leads to a steadier current trajectory that does not activate the lower constraint bound. As can be seen in Fig.~\ref{fig:u_y_no_noise}, this behavior is achieved at the expense of  braking performance.

\begin{figure}[t]
    \centering
    \input{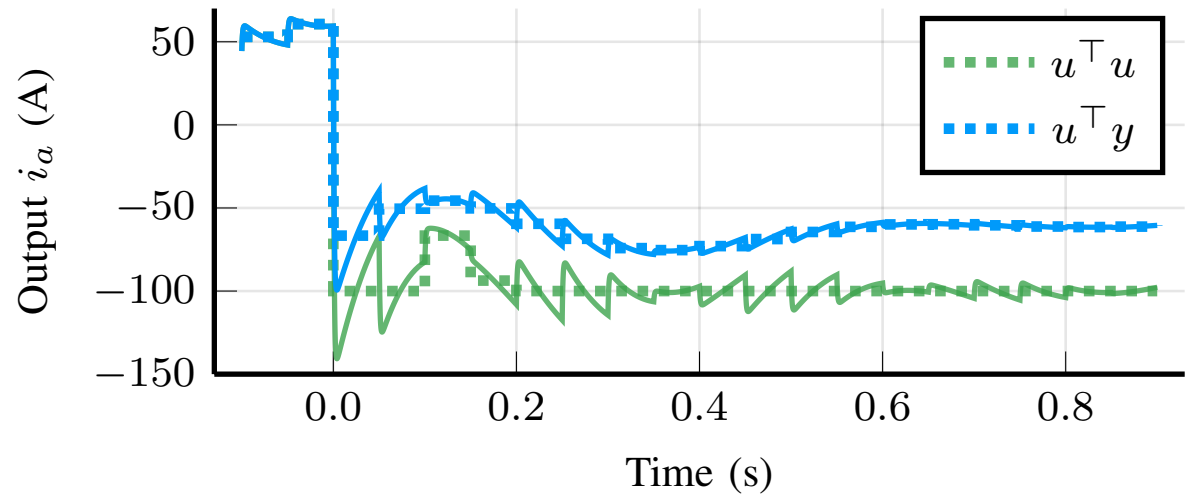}
    \caption{Close-up comparison of the closed-loop output current $i_a$ for the two controllers with OCP stage costs $u^\top u+ y^\top Q y$ (green) and $u^\top y+ y^\top Q y$ (blue). Both OCPs here are formulated in the reduced coordinates $g_0\oplus \operatorname{im}(N_\mathrm{p})$ as in \eqref{eq:convex_OCP}. Dashed lines show currents discretized via \eqref{eq:sampled_y} and solid lines show the corresponding continuous-time currents.}
    \label{fig:ia_close_up}
\end{figure}

Note that due to the discretization, the OCPs are only able to constrain the output averages over sampling time intervals and they do not oversee the actual continuous-time output trajectories. Fig.~\ref{fig:ia_close_up} depicts the current behavior achieved by the data-driven controllers. While the controller solving OCPs with the term $u^\top u$ keeps the average current near $-100~\si{\ampere}$, the continuous-time current ripples cross this mark multiple times. This happens because the $50~\si{\milli\second}$ sampling time prevents controllers from tracking the significantly faster current dynamics.

In Tab.~\ref{tab:solver_performance}, we examine the average and standard deviation (std) times needed to solve the $100$ data-driven OCPs within the first $100 \cdot 50~\si{\milli\second} = 5~\si{\second}$ after initiating the braking shown in Fig.~\ref{fig:u_y_no_noise}. By reformulating \eqref{eq:OCP} in the reduced coordinates \eqref{eq:column_selector} we can guarantee the convexity of its objective, which allows us to use the faster \texttt{DAQP.jl} solver tailored to convex QPs as opposed to the general-purpose \texttt{Ipopt.jl}. Thus, we achieve a significant speed-up, with average computation times well below the sampling time. Notably, the first two OCPs from Tab.~\ref{tab:solver_performance} appear equivalent. In simulation, the maximum absolute difference in the produced control inputs (blue solid line and dashed orange line in Fig.~\ref{fig:u_y_no_noise}) is $\approx 3.645 \cdot 10^{-7}~\si{\volt}$, which confirms Proposition~\ref{pro:convex_reformulation}.

\begin{table}[t]
    \centering
    \caption{Closed-loop statistics over the first $100$ time steps.}
    \label{tab:solver_performance}
    \begin{tabular}{lccc}
        \toprule
        \multirow{2}{*}{\thead[l]{Stage cost ($-y^\top Q y$),\\ column selector space}}
            & \multicolumn{2}{c}{OCP solve time (\si{\second})} & \multirow{2}{*}{\thead[l]{$E_\mathrm{diss}$ \\ (\si{\kilo\joule})}} \\
        \cmidrule(lr){2-3}
            & Mean & Std & \\
        \midrule
        $u^\top y$, $\mathbb R^{n_g}$ & $7.808$ & $12.753$ & $1.948$ \\
        $u^\top y$, $g_0 \oplus \operatorname{im}(N_\mathrm{p})$ & $1.237 \cdot 10^{-3}$ & $0.882\cdot 10^{-3}$ & $1.948$\\
        $u^\top u$, $g_0 \oplus \operatorname{im}(N_\mathrm{p})$ & $1.117\cdot 10^{-3}$ & $0.775\cdot 10^{-3}$ & $2.802$\\
        \bottomrule
    \end{tabular}
\end{table}

\subsection{Output Measurement Noise}
In applications, output averages \eqref{eq:sampled_y} are subject to measurement errors, in particular whenever they are approximated via \eqref{eq:sampled_y_approx}. Here, we assume the latter sampling scheme with $T_s^\prime = T_s/10 = 5~\si{\milli\second}$. Furthermore, we corrupt each individual output measurement in \eqref{eq:sampled_y_approx} using additive zero-mean Gaussian noise with standard deviation equal to $1\%$ of the signal magnitude. 

To this end, we reset $Q\doteq \operatorname{diag}([0.8\;\;5]) \cdot 10^{-4}$. We investigate how the passivity-consistent regularization in our modified OCP \eqref{eq:practical_OCP} compares to \eqref{eq:convex_OCP} equipped with a standard quadratic regularization $\lambda_\mathrm{q} \lVert \eta\rVert_2^2$, $\lambda_\mathrm{q} \in \mathbb R_{>0}$. To induce active output constraints, we limit the current $i_a$ to $[-60, 60]~\si{\ampere}$. In Tab.~\ref{tab:noisy_performance}, we judge the effectiveness of the considered regularizations in closed-loop based on the number of (discrete-time) output constraint violations, total dissipated energy, and braking performance observed within the first $100 \cdot 50~\si{\milli\second} = 5~\si{\second}$ since initiating the braking. Since the measurement noise is stochastic, we calculate the mean and standard deviation for all of the mentioned quantities using Monte Carlo simulations. In each Monte Carlo MPC run, data for the Hankel matrices are re-collected for the same input signal but with re-sampled output measurement noise realizations. For each OCP, the output measurements $y_{\text p,k}$ in the immediate past are also subject to the noise, whose realizations are re-sampled with each Monte Carlo run.

As can be seen from Tab.~\ref{tab:noisy_performance},  using regularization \eqref{eq:H_practical}  with $\lambda=1$, we are able to achieve the fewest number of output current violations both in terms of mean and standard deviation compared to \eqref{eq:convex_OCP} with quadratic regularization $\lambda_\mathrm{q} \lVert \eta\rVert_2^2$ for a wide range of $\lambda_\mathrm{q}$. Similarly, on average, the controller based on the OCP \eqref{eq:practical_OCP} achieves the smallest dissipated energy. We attribute this performance to the penalization of the components of the input-output behavior that are inconsistent with the passivity of the system. In contrast, the quadratic term $\lambda_\mathrm{q} \lVert \eta\rVert_2^2$ penalizes all input-output predictions merely based on their magnitude without specifically filtering out the passivity-inconsistent part.

\begin{table}[t]
    \centering
    \caption{Closed-loop statistics over the first $100$ time steps and $100$ Monte Carlo iterations.}
    \label{tab:noisy_performance}
    \small
    \begin{tabular}{lcccccc}
        \toprule
        \multirow{2}{*}{$\lambda_\mathrm{q}$}
            & \multicolumn{2}{c}{\# Violations}
            & \multicolumn{2}{c}{$E_\mathrm{diss}$ (\si{\joule})}
            & \multicolumn{2}{c}{$\sum y^\top Q y$} \\
        \cmidrule(lr){2-3} \cmidrule(lr){4-5} \cmidrule(lr){6-7}
            & Mean & Std & Mean & Std & Mean & Std \\
        \midrule
        $0.01$ & 27.81 & 44.45 & 3847.7 & 3144.5 & 3136.2 & 1278.6 \\
        $1$ & 1.94 & 9.46 & 1763.1 & 138.1 & 1744.7 & 314.2 \\
        $10$ & 3.88 & 14.85 & 1973.4 & 62.1 & 1371.7 & 47.6 \\
        $100$ & 27.99 & 14.92 & 2073.3 & 10.8 & 1306.0 & 6.8 \\
        \midrule
        $\lambda = 1$ & 0.25 & 1.38 & 1658.4 & 105.2 & 1887.2 & 254.1 \\
        \bottomrule
    \end{tabular}
\end{table}

\section{Conclusions}~\label{sec:con}
We present a data-driven predictive control framework for unknown  passive LTI systems. We employ passivity-preserving sampling to retain the continuous-time energy balance in the sampled input-output data. In the control objective, we replace conventional quadratic input regularization with the passive supply rate to promote energy extraction. We exploit passivity-induced positive semidefiniteness to reformulate the resulting bilinear objective as an equivalent convex quadratic program. Together with the objective design, terminal equality constraints ensure recursive feasibility and asymptotic output convergence. Numerical simulations demonstrate the closed-loop performance and computational benefits of the convex reformulation. Future work will explore extensions to more general dissipative and nonlinear  systems.
\subsection*{Acknowledgments}
The authors would like to thank Prof. Paolo Rapisarda for very helpful discussions regarding this paper.

The authors acknowledge the use of OpenAI's ChatGPT
for language editing and improving the clarity of the
manuscript, and Anthropic's Claude Code for assistance
with coding. All technical content, mathematical
results, and conclusions were developed and verified
by the authors.

\bibliographystyle{IEEEtran}
\bibliography{ALL_GP, oleksii}

@Article{Willems2005,
  author    = {Willems, J. C. and Rapisarda, P. and Markovsky, I. and De Moor, B. L. M.},
  journal   = {Systems \& Control Letters},
  title     = {A note on persistency of excitation},
  year      = {2005},
  number    = {4},
  pages     = {325--329},
  volume    = {54},
  doi       = {10.1016/j.sysconle.2004.09.003},
  publisher = {Elsevier},
}

@Article{Garc_s_Ruiz_2024,
  author    = {Garcés-Ruiz, Alejandro and Gil González, Walter Julián},
  journal   = {Ingeniería},
  title     = {Passivity-Based Model-Predictive Control for the Permanent Magnet Synchronous Machine},
  year      = {2024},
  issn      = {0121-750X},
  month     = Sept,
  number    = {3},
  pages     = {e22162},
  volume    = {29},
  doi       = {10.14483/23448393.22162},
  publisher = {Universidad Distrital Francisco Jose de Caldas},
}

@Article{Falugi_2014,
  author    = {Falugi, P.},
  journal   = {IFAC Proceedings Volumes},
  title     = {Model predictive control: a passive scheme},
  year      = {2014},
  issn      = {1474-6670},
  number    = {3},
  pages     = {1017--1022},
  volume    = {47},
  doi       = {10.3182/20140824-6-za-1003.02165},
  publisher = {Elsevier BV},
}

@InProceedings{Coulson2019,
  author       = {Coulson, Jeremy and Lygeros, John and D{\"o}rfler, Florian},
  booktitle    = {2019 18th European Control Conference (ECC)},
  title        = {Data-enabled predictive control: In the shallows of the {DeePC}},
  year         = {2019},
  organization = {IEEE},
  pages        = {307--312},
}

@Article{Berberich2020,
  author    = {Berberich, Julian and K{\"o}hler, Johannes and M{\"u}ller, Matthias A. and Allg{\"o}wer, Frank},
  journal   = {IEEE Transactions on Automatic Control},
  title     = {Data-driven model predictive control with stability and robustness guarantees},
  year      = {2020},
  number    = {4},
  pages     = {1702--1717},
  volume    = {66},
  publisher = {IEEE},
}

@Article{Markovsky2021,
  author    = {Markovsky, I. and D{\"o}rfler, F.},
  journal   = {Annual Reviews in Control},
  title     = {Behavioral systems theory in data-driven analysis, signal processing, and control},
  year      = {2021},
  issn      = {1367-5788},
  pages     = {42--64},
  volume    = {52},
  doi       = {10.1016/j.arcontrol.2021.09.005},
  publisher = {Elsevier BV},
}

@Article{Markovsky2008,
  author    = {Markovsky, Ivan and Rapisarda, Paolo},
  journal   = {International Journal of Control},
  title     = {Data-driven simulation and control},
  year      = {2008},
  number    = {12},
  pages     = {1946--1959},
  volume    = {81},
  doi       = {10.1080/00207170801942170},
  publisher = {Taylor \& Francis},
}

@Book{Schaft2017,
  author    = {van der Schaft, Arjan},
  publisher = {Springer International Publishing},
  title     = {{L2-Gain and Passivity Techniques in Nonlinear Control}},
  year      = {2017},
  isbn      = {9783319499925},
  doi       = {10.1007/978-3-319-49992-5},
  issn      = {2197-7119},
  journal   = {Communications and Control Engineering},
}

@Article{Willems1972,
  author    = {Willems, Jan C.},
  journal   = {Archive for Rational Mechanics and Analysis},
  title     = {{Dissipative dynamical systems part I: General theory}},
  year      = {1972},
  issn      = {1432-0673},
  number    = {5},
  pages     = {321--351},
  volume    = {45},
  doi       = {10.1007/BF00276493},
  publisher = {Springer Science and Business Media LLC},
}

@Article{Schaller2021,
  author    = {Schaller, Manuel and Philipp, Friedrich and Faulwasser, Timm and Worthmann, Karl and Maschke, Bernhard},
  journal   = {European Journal of Control},
  title     = {Control of port-{H}amiltonian systems with minimal energy supply},
  year      = {2021},
  issn      = {0947-3580},
  month     = nov,
  pages     = {33--40},
  volume    = {62},
  doi       = {https://doi.org/10.1016/j.ejcon.2021.06.017},
  publisher = {Elsevier BV},
}

@Article{Stramigioli2005,
  author   = {Stramigioli, S. and Secchi, C. and van der Schaft, A. J. and Fantuzzi, C.},
  journal  = {IEEE Transactions on Robotics},
  title    = {Sampled data systems passivity and discrete port-{H}amiltonian systems},
  year     = {2005},
  number   = {4},
  pages    = {574-587},
  volume   = {21},
  doi      = {10.1109/TRO.2004.842330},
}

@InCollection{Raff2007,
  author    = {Raff, Tobias and Ebenbauer, Christian and Allg{\"o}wer, Prank},
  booktitle = {Assessment and future directions of nonlinear model predictive control},
  publisher = {Springer},
  title     = {Nonlinear model predictive control: A passivity-based approach},
  year      = {2007},
  pages     = {151--162},
}

@Article{Romer2019,
  author   = {Romer, Anne and Berberich, Julian and Köhler, Johannes and Allgöwer, Frank},
  journal  = {IEEE Control Syst. Lett.},
  title    = {One-shot verification of dissipativity properties from input–output data},
  year     = {2019},
  number   = {3},
  pages    = {709-714},
  volume   = {3},
  doi      = {10.1109/LCSYS.2019.2917162},
}

@Article{Maupong2017,
  author    = {Maupong, T. M. and Mayo-Maldonado, J. C. and Rapisarda, P.},
  journal   = {IFAC-PapersOnLine},
  title     = {On {Lyapunov} functions and data-driven dissipativity},
  year      = {2017},
  issn      = {2405-8963},
  month     = July,
  number    = {1},
  pages     = {7783--7788},
  volume    = {50},
  doi       = {https://doi.org/10.1016/j.ifacol.2017.08.1052},
  publisher = {Elsevier BV},
}

@Article{LorancaCoutino2022,
  author   = {Loranca-Coutiño, Javier and Mayo-Maldonado, Jonathan C. and Escobar, Gerardo and Maupong, Thabiso M. and Valdez-Resendiz, Jesus E. and Rosas-Caro, Julio C.},
  journal  = {IEEE Transactions on Industrial Electronics},
  title    = {Data-driven passivity-based control design for modular {DC} microgrids},
  year     = {2022},
  number   = {3},
  pages    = {2545-2556},
  volume   = {69},
  doi      = {10.1109/TIE.2021.3065615},
}

@Article{Jiang1993,
  author    = {Jiang, Jin},
  journal   = {Journal of the Franklin Institute},
  title     = {Preservations of positive realness under discretizations},
  year      = {1993},
  issn      = {0016-0032},
  month     = July,
  number    = {4},
  pages     = {721--734},
  volume    = {330},
  doi       = {https://doi.org/10.1016/0016-0032(93)90050-5},
  publisher = {Elsevier BV},
}

@Book{Chen1999,
  author    = {Chi-Tsong Chen},
  publisher = {Oxford University Press},
  title     = {Linear system theory and design},
  year      = {1999},
  address   = {New York},
  edition   = {3},
  isbn      = {0-19-511777-8},
  pagetotal = {334},
  ppn_gvk   = {1659143500},
}

@Article{Angeli2012,
  author   = {Angeli, David and Amrit, Rishi and Rawlings, James B.},
  journal  = {IEEE Transactions on Automatic Control},
  title    = {On average performance and stability of economic model predictive control},
  year     = {2012},
  number   = {7},
  pages    = {1615--1626},
  volume   = {57},
  doi      = {10.1109/TAC.2011.2179349},
}

@Article{Faulwasser2018,
  author    = {Faulwasser, Timm and Grüne, Lars and Müller, Matthias A.},
  journal   = {Foundations and Trends® in Systems and Control},
  title     = {Economic nonlinear model predictive control},
  year      = {2018},
  issn      = {2325-6826},
  month     = Jan,
  number    = {1},
  pages     = {1--98},
  volume    = {5},
  doi       = {10.1561/2600000014},
  publisher = {Emerald},
}

@Article{Zanon2014,
  author  = {Mario Zanon and Sébastien Gros and Moritz Diehl},
  journal = {Journal of Process Control},
  title   = {Indefinite linear {MPC} and approximated economic {MPC} for nonlinear systems},
  year    = {2014},
  issn    = {0959-1524},
  number  = {8},
  pages   = {1273--1281},
  volume  = {24},
  doi     = {10.1016/j.jprocont.2014.04.023},
}

@Article{Sen2023,
  author    = {Şen, Gökçen Devlet and Schaller, Manuel and Worthmann, Karl},
  journal   = {PAMM},
  title     = {Stage‐cost design for optimal and model predictive control of linear port‐{H}amiltonian systems: Energy efficiency and robustness},
  year      = {2023},
  issn      = {1617-7061},
  month     = Oct,
  number    = {4},
  volume    = {23},
  doi       = {10.1002/pamm.202300296},
  publisher = {Wiley},
}

@Misc{Singh2026,
  author    = {Singh, Vaibhav Kumar and Schaller, Manuel and Faulwasser, Timm and Worthmann, Karl},
  title     = {Energy-optimal predictive control of discrete-time port-{H}amiltonian systems: Closed-loop practical stability},
  year      = {2026},
  copyright = {Creative Commons Attribution Non Commercial No Derivatives 4.0 International},
  doi       = {10.48550/ARXIV.2609.25920},
  publisher = {arXiv},
}

@Article{Sarkar2025,
  author   = {Sarkar, Arijit and Kumar Singh, Vaibhav and Schaller, Manuel and Worthmann, Karl},
  journal  = {IEEE Control Systems Letters},
  title    = {Energy-optimal control of discrete-time port-{H}amiltonian systems},
  year     = {2025},
  pages    = {1526-1531},
  volume   = {9},
  doi      = {10.1109/LCSYS.2025.3579391},
}

@article{arnstrom2022dual,
  author={Arnström, Daniel and Bemporad, Alberto and Axehill, Daniel},
  journal={IEEE Transactions on Automatic Control},
  title={A Dual Active-Set Solver for Embedded Quadratic Programming Using Recursive {LDL}$^{T}$ Updates},
  year={2022},
  volume={67},
  number={8},
  pages={4362-4369},
  doi={10.1109/TAC.2022.3176430}
}

@Article{ipopt,
  author    = {W{\"a}chter, Andreas and Biegler, Lorenz T.},
  journal   = {Mathematical Programming},
  title     = {On the implementation of an interior-point filter line-search algorithm for large-scale nonlinear programming},
  year      = {2006},
  number    = {1},
  pages     = {25--57},
  volume    = {106},
  publisher = {Springer},
}
\end{document}